\documentclass[12pt]{iopart}

\usepackage{iopams} 
\expandafter\let\csname equation*\endcsname\relax
\expandafter\let\csname endequation*\endcsname\relax 
\usepackage{amsmath}  
\usepackage{graphicx}
\usepackage{subfigure}
\input epsf
\epsfclipon
\usepackage{epstopdf}

\begin{document}

\title[Exact solution of the isotropic majority-vote model on complete graphs] {Exact solution of the isotropic majority-vote model on complete graphs}

\author{Agata Fronczak and Piotr Fronczak}

\address{Faculty of Physics, Warsaw University of Technology, Koszykowa 75,\\ PL-00-662 Warsaw, Poland}
\ead{agatka@if.pw.edu.pl, fronczak@if.pw.edu.pl}
\vspace{10pt}

\begin{abstract}
The isotropic majority-vote (MV) model which, apart from the one-dimensional case, is thought to be non-equilibrium and violating the detailed balance condition. We show that this is not true, when the model is defined on a complete graph. In the stationary regime, the MV model on a fully connected graph fulfills the detailed balance. We derive the exact expression for the  probability distribution of finding the system in a given spin configuration. We show that it only depends on the absolute value of magnetization. Our theoretical predictions are validated by numerical simulations.
\end{abstract}


\vspace{2pc}

\section{Introduction}\label{SectIntro}

The isotropic majority-vote (MV) model for opinion dynamics is a well-known nonequilibrium spin model which has been studied by many researchers, see eg.~\cite{1992_JStatPhysOliveira, 2005_PREPereira, 2007_PREKwak, 2008_PREYang, 2011_PhysASantos, 2012_JStatMechTilles, 2012_PRELara, 2014_PRELara, 2015_PREChen, 2015_APPAGradowski, 2016_EntropyLima}. One of the reasons, why physicists became interested in the model, is its critical behavior. The model was originally introduced in 1992, in Ref.~\cite{1992_JStatPhysOliveira}, where, for the two-dimensional square lattice, it was shown to have a continuous phase transition with the same static critical exponents as the 2D Ising model. This result was an important confirmation of an earlier hypothesis according to which nonequilibrium models with up-down symmetry and spin-flip dynamics fall within the universality class of the equilibrium Ising model \cite{1985_PRLGrinstein, 1993_JPhysAOliveira}. Since then, there have been many numerical studies aimed at final approval or rejection of that hypothesis for different models. In particular, a number of Monte Carlo simulations of the MV model on regular lattices and random networks have been carried out providing contradictory conclusions (cf.~eg.~\cite{2008_PREYang} and \cite{2012_PRELara}) and leaving unsolved the problem of its universality classes. For example, it is still controversial whether the upper critical dimension of the MV model on $d$-dimensional hypercubic lattices is $4$ (as in the Ising model) or $6$ (as suggested in \cite{2008_PREYang}). Moreover, mean-field critical exponents for the MV model have not yet been exactly determined, although their numerical values seem to agree with the results known for the Ising model.

This contribution is the first of two, in which we refer to the last point among those listed at the end of the previous paragraph. Here, we present exact solution of the isotropic majority-vote model on the complete graph of $N$ nodes. We provide the probability distribution, $P(\Omega)$, of finding the system in a certain microstate $\Omega=(\sigma_1,\sigma_2,\dots,\sigma_N)$, where $\sigma_i=\pm 1$ denotes the spin variable associated to the site $i$ and $N$ is the total number of sites. We show that the probability depends only on the absolute value of magnetization: $P(\Omega)\propto \sqrt{(1-q)/q}^{|M(\Omega)|}$, where $M(\Omega)=\sum_{i=1}^N\sigma_i$ and $q$ is the standard noise parameter of the majority-vote model (see Eq.~(\ref{rate0}) for its formal definition). Given the result, we also show that the MV model on the complete graph fulfills the detailed-balance condition: just as the one-dimensional MV model \cite{2015_bookTome} and unlike the model on the two-dimensional square lattice \cite{1992_JStatPhysOliveira}. In the next paper, we study critical properties of the model and determine exact values of its mean-field critical exponents. 


\section{A brief state-of-the-art on the majority-vote model}\label{SectBackground}

The isotropic majority-vote model is a spin model in which to each site $i$ a spin variable $\sigma_i=\pm 1$ is assigned. In the rest of this paper, the global configuration of the system is denoted by 
\begin{equation}\label{micro0}
\Omega=(\sigma_1,\sigma_2,\dots,\sigma_i,\dots \sigma_N),
\end{equation}
where $N$ is the total number of sites. The dynamics of the model is such that, at any time step, only one site has its spin modified. Let us assume that the considered spin is $\sigma_i$. The transition rate from a configuration $\Omega$ to another configuration
\begin{equation}\label{microi}
\Omega_{i}=(\sigma_1,\sigma_2,\dots,-\sigma_i,\dots \sigma_N),
\end{equation}
which only differs from $\Omega$ by the sign of the $i$th spin variable, is given by
\begin{equation}\label{rate0}
w_i(\Omega)=\frac{1}{2}[1-(1-2q)\sigma_iS_i],
\end{equation}
where $S_i$ takes one of three values 
\begin{equation}\label{defSi}
S_i=S(m_i)= \left\{ \begin{array}{lcl}
+1 & \mbox{for } &m_i>0 \\
\;\;\;0 & \mbox{for } & m_i=0\\
-1 & \mbox{for } & m_i<0
\end{array}\right.,
\end{equation}
depending on the local magnetization,
\begin{equation}\label{defmi}
m_i=\sum_{\langle i,j\rangle}\sigma_j,
\end{equation} 
of the nearest neighborhood of $\sigma_i$.

The role of the noise parameter $q$ can be easily deduced from Eq.~(\ref{rate0}). In the case, when, in the initial configuration $\Omega$, the sign of the variable $\sigma_i$ is consistent with the sign of its neighborhood, i.e. $\sigma_iS_i=+1$, the transition rate, which is proportional to the probability that $\sigma_i$ changes the sign to the opposite, is equal to $w_{i}(\Omega)=q$. Otherwise, when $\sigma_iS_i=-1$ one has $w_{i}(\Omega)=1-q$. Thus, the transition rate from a spin configuration $\Omega$ to $\Omega_i$ is $1-q$ if the flipping follows the majority rule among the nearest neighbors of $i$, and $q$ if it does not. The case when $S_i=0$ corresponds to $w_i(\Omega)=\frac{1}{2}$, which means that the chosen site takes either sign with equal probability. From the above discussion, it is clear that the parameter $q$ in the MV model can be interpreted as an effective temperature. However, the meaning of $q$ is much richer. In particular, when $0\leq q<\frac{1}{2}$ the majority rule mimics a kind of ferromagnetic coupling in the system, while $\frac{1}{2}<q\leq 1$ corresponds to anti-ferromagnetic coupling. Finally, when $q=\frac{1}{2}$ the model behaves like a typical paramagnet. 
 
The time evolution of the majority-vote model is governed by the master equation:
\begin{equation}\label{masterEq0}
\frac{d}{dt}P(\Omega,t)=\sum_{i}\left[w_i(\Omega)P(\Omega,t)- w_i(\Omega_i)P(\Omega_i,t)\right],
\end{equation}
where $P(\Omega,t)$ is the  time-dependent microstate distribution, i.e. the probability of occurrence of configuration $\Omega$ at time $t$, and $w_i(\Omega)$ is the transition rate from $\Omega$ to $\Omega_i$, which is given by Eq.~(\ref{rate0}). In the stationary regime, when
\begin{equation}\label{masterEq1}
\frac{d}{dt}P(\Omega,t)=0,\mbox{\;\;\;\;\;i.e.\;\;\;\;\;} P(\Omega,t)\equiv P(\Omega),
\end{equation}
Eq.~(\ref{masterEq0}) simplifies to the following balance equation:
\begin{equation}\label{masterEq2}
\sum_{i}\left[w_i(\Omega)P(\Omega)- w_i(\Omega_i)P(\Omega_i)\right]=0,
\end{equation}
which, in general and differently than it is in the Ising model, can not be further simplified to the detailed balance condition:
\begin{equation}\label{masterEq3}
w_i(\Omega)P(\Omega)-w_i(\Omega_i)P(\Omega_i)\neq 0.
\end{equation}

With regard to the majority-vote model, the only known exception to Eq.~(\ref{masterEq3}) is the one-dimensional chain of spins, which is equivalent to the 1D Glauber model, whose dynamics may be interpreted as a dynamics for the 1D Ising model (see \cite{2015_bookTome}, Chap.~11). In other words, in the stationary regime, the one-dimensional MV model is equivalent to the 1D Ising model. Therefore, its stationary distribution $P(\Omega)$ is the Boltzmann-Gibbs distribution. In higher dimensions, one believes that the MV model does not show microscopic reversibility which underlies the detailed balance, although, in a similar way to what happens in the microscopically reversible Ising model, the MV model exhibits continuous phase transitions. (The lack of microscopic reversibility in the square-lattice MV model is simply shown in Ref.~\cite{2015_bookTome}, Chap.~12.6.) For example, in the stationary regime, for small values of the parameter $q$, the square-lattice MV model presents a ferromagnetic phase characterized by the presence of a majority of spins with the same sign. Above the critical value $q_c=0.075\pm 0.01$ \cite{1992_JStatPhysOliveira} of the noise parameter, the model presents a paramagnetic state with equal, in the average, number of spins with distinct signs. Furthermore, it is known that the square-lattice MV model falls into the same universality class as the 2D Ising model. Recent numerical simulations indicate that these findings may also be true in higher dimensions, for $d\geq 3$ \cite{2012_PRELara}. On the other hand, one still lacks strict theoretical results concerning, in particular, the mean field critical behavior of the model.  

In the next section, starting with the balance equation, Eq.~(\ref{masterEq2}), we find recurrence relations for the probability, $P(\Omega)$, of finding the MV model on a complete graph in a given spin configuration, $\Omega$. Then, we derive exact formulas for $P(\Omega)$ and $P(M)$, with the latter being the probability that the system has a magnetization equal to $M$. We also show that the MV model on complete graphs fulfills the detailed balance condition. Our theoretical predictions are confirmed by numerical simulations.

\section{Majority-vote model on the complete graph}

\subsection{Probability of a configuration in the stationary regime}\label{Theory}

Let us note that, in the case of a complete graph, the local magnetization, Eq.~(\ref{defmi}), of each node is
\begin{equation}\label{Kmi}
m_i=\sum_{j\neq i}\sigma_j=M-\sigma_i,
\end{equation}
where 
\begin{equation}\label{KM}
M=\sum_{j}\sigma_{j},
\end{equation}
is the total magnetization of the system in a spin configuration $\Omega$. Later in the text, if not explicitly stated otherwise, quantities such as magnetization, $M$, the number of positive spins, $N_{\!+}$, and the spin variable, $\sigma_i$, always refer to the configuration $\Omega$. Accordingly, if we want to emphasize that these variables refer to another configuration, e.g. $\Omega_i$, we write it in the following way: $M(\Omega_i)$, $N_{\!_+}(\Omega_i)$, and $\sigma_i(\Omega_i)$. 

Given Eq.~(\ref{Kmi}), transition rates in the balance equation, Eq.~(\ref{masterEq2}), can be written as follows:
\begin{equation}\label{rate1a}
w_i(\Omega)=\frac{1}{2}[1-(1-2q)\sigma_iS(M-\sigma_i)],
\end{equation}
\begin{equation}\label{rate1b}
w_i(\Omega_i)=\frac{1}{2}[1+(1-2q)\sigma_iS(M-\sigma_i)],
\end{equation}
where the function $S(m_i)$ is that defined in Eq.~(\ref{defSi}) and $m_i(\Omega)=m_i(\Omega_i)=M-\sigma_i$. These rates only depend on the magnetization, $M$, and on the sign of the spin variable, $\sigma_i$. Therefore, in a compete graph, since all the spins with the same sign are equivalent, the balance equation (cf. with Eq.~(\ref{masterEq2})) 
\begin{equation}\label{masterEq4}
P(\Omega)\sum_iw_i(\Omega)=\sum_iP(\Omega_i)w_i(\Omega_i),
\end{equation}
gets the following form:
\begin{equation}\label{masterEq5}
N_{\!_+}[1-(1\!-\!2q)S(M\!-\!1)]P(\Omega)+N_{\!_-}[1+(1\!-\!2q)S(M\!+\!1)]P(\Omega)=
\end{equation} 
\begin{equation}\nonumber
N_{\!_+}[1+(1\!-\!2q)S(M\!-\!1)]P(\Omega_{_-})+N_{\!_-}[1-(1\!-\!2q)S(M\!+\!1)]P(\Omega_{_+}),\;\;\;
\end{equation}
where $\Omega_{_-}$ and $\Omega_{_+}$ stand for the configurations $\Omega_i$ such that $\sigma_i(\Omega_{_-})=-\sigma_i(\Omega)=-1$ and  $\sigma_i(\Omega_{_+})=-\sigma_i(\Omega)=+1$, respectively. 

To proceed with the analysis of Eq.~(\ref{masterEq5}), one must assume that the system size, $N$, is even or odd. At the beginning, we assume that $N$ is even. Therefore, the magnetization
\begin{equation}\label{defM}
M=2N_{\!_+}-N,
\end{equation}
is also even (positive or negative) or zero, and the balance equation, Eq.~(\ref{masterEq5}), splits into three cases depending on the sign of $M$. Thus, we have 
\begin{equation}\label{masterEq6a}
\mbox{for   }M\leq\!-2\!:\;\;\; N_{\!_+}(1\!-\!q)P(\Omega)+N_{\!_-}qP(\Omega)=N_{\!_+}qP(\Omega_{_-})+N_{\!_-}(1\!-\!q)P(\Omega_{_+}),
\end{equation}
\begin{equation}\label{masterEq6b}
\mbox{for   } M=\!0\!:\;\;\;\;\;\;N_{\!_+}(1\!-\!q)P(\Omega)+N_{\!_-}(1\!-\!q)P(\Omega)=N_{\!_+}qP(\Omega_{_-})+N_{\!_-}qP(\Omega_{_+}),
\end{equation}
\begin{equation}\label{masterEq6c}
\mbox{for   } M\geq\!+2\!:\;\;\;N_{\!_+}qP(\Omega)+N_{\!_-}(1\!-\!q)P(\Omega)=N_{\!_+}(1\!-\!q)P(\Omega_{_-})+N_{\!_-}qP(\Omega_{_+}).
\end{equation}

The above equations can be significantly simplified if one assumes that the probability of a configuration, $P(\Omega)$, only depends on the number of positive spins, $N_{\!_+}(\Omega)$, i.e.
\begin{equation}\label{assum0}
P(\Omega)\equiv f(N_{\!_+}(\Omega))=f(N_{\!_+}), 
\end{equation} 
and correspondingly
\begin{equation}\label{assum1}
\begin{array}{rcl}
P(\Omega_{_+})\!&\!\equiv\!&\!f(N_{\!_+}(\Omega_{_+}))=f(N_{\!_+}\!\!+\!1),\\
P(\Omega_{_-})\!&\!\equiv\!&\!f(N_{\!_+}(\Omega_{_-}))=f(N_{\!_+}\!\!-\!1).
\end{array}
\end{equation} 
The assumptions provided by Eqs.~(\ref{assum0})-(\ref{assum1}) are reasonable due to the symmetry of the system, in which all the spins with the same sign are equivalent. Also, they naturally arise from the balance equations, Eqs.~(\ref{masterEq6a})-(\ref{masterEq6c}), in which transition rates between different configurations only depend on $N_{\!_+}$. According to these assumptions, one gets the following recurrence relations for the dummy function $f(N_{\!_+})$:
\begin{equation}\label{masterEq7a}
\mbox{for   }N_{\!_+}<\!\frac{N}{2}\!:\;\;\;\; 
N_{\!_+}(1\!-\!q)f(N_{\!_+})+N_{\!_-}qf(N_{\!_+})=N_{\!_+}qf(N_{\!_+}\!\!-\!1)+N_{\!_-}(1\!-\!q)f(N_{\!_+}\!\!+\!1),
\end{equation}
\begin{equation}\label{masterEq7b}
\mbox{for   } N_{\!_+}=\!\frac{N}{2}\!:\;\;\;\;
N_{\!_+}(1\!-\!q)f(N_{\!_+})+N_{\!_-}(1\!-\!q)f(N_{\!_+})=N_{\!_+}qf(N_{\!_+}\!\!-\!1)+N_{\!_-}qf(N_{\!_+}\!\!+\!1),
\end{equation}
\begin{equation}\label{masterEq7c}
\mbox{for   } N_{\!_+}>\!\frac{N}{2}\!:\;\;\;\;
N_{\!_+}qf(N_{\!_+})+N_{\!_-}(1\!-\!q)f(N_{\!_+})=N_{\!_+}(1\!-\!q)f(N_{\!_+}\!\!-\!1)+N_{\!_-}qf(N_{\!_+}\!\!+\!1).
\end{equation}

At first glance the above relations seem quite complicated, but in fact they have a fairly simple structure. In particular, it is easy to see that Eq.~(\ref{masterEq7a}), which is valid for $N_{\!_+}=0,1,2,\dots,\frac{N}{2}-1$, can be written in the following form:  
\begin{equation}\label{masterEq8a}
N_{\!_+}\;F(N_{\!_+}\!\!-\!1)=N_{\!_-}\;F(N_{\!_+}),
\end{equation}
where
\begin{equation}\label{masterEq9a}
F(N_{\!_+})=(1\!-\!q)f(N_{\!_+}\!\!+\!1)-qf(N_{\!_+}).
\end{equation}
When examining Eq.~(\ref{masterEq8a}) for $N_{\!_+}=0$ one gets $F(0)\!=\!0$. Then, using $F(0)\!=\!0$ in the same equation for $N_{\!_+}=1$, one gets $F(1)\!=\!0$. In a similar way, one can show that for each value of $N_{\!_+}<\frac{N}{2}$, one has $F(N_{\!_+})\!=\!0$, i.e.  
\begin{equation}\label{masterEq10a}
\mbox{for}\;\;N_{\!_+}=0,1,2,\dots,\frac{N}{2}\!-\!1:\hspace{1.5cm} f(N_{\!_+})=\frac{1\!-\!q}{q}f(N_{\!_+}\!\!+\!1).
\end{equation}
Hence, for the subsequent values of $N_{\!_+}$ one gets:
\begin{equation}\label{masterEq11a}
f\!\left(\frac{N}{2}\!-\!1\right)= f\!\left(\frac{N}{2}\right)\left(\frac{1\!-\!q}{q}\right)^{\;}\!,
\end{equation}
\begin{equation}\label{masterEq11b}
f\!\left(\frac{N}{2}\!-\!2\right)=f\!\left(\frac{N}{2}\right)\left(\frac{1\!-\!q}{q}\right)^{\!2}\!,
\end{equation}
\begin{equation}\nonumber
\dots
\end{equation}
and finally, for $N_{\!_+}=\frac{N}{2}+\frac{M}{2}$, where $M<0$ (see Eq.~(\ref{defM})), the dummy function $f(N_{\!_+})$ is given by:
\begin{equation}\label{masterEq12a}
f\!\left(N_{\!_+}\right)= f\!\left(\frac{N}{2}\right)\left(\frac{1\!-\!q}{q}\right)^{\!-\frac{M}{2}} \!\!.
\end{equation}
In a similar way, from Eq.~(\ref{masterEq7c}), one can show that (cf.~with Eq.~\ref{masterEq10a})
\begin{equation}\label{masterEq10b}
\mbox{for}\;\;N_{\!_+}=\frac{N}{2}\!+\!1,\frac{N}{2}\!+\!2,\dots,N:\hspace{1.5cm} f(N_{\!_+}\!)=\frac{1\!-\!q}{q}\;f(N_{\!_+}\!\!-\!1),
\end{equation}
and hence for $N_{\!_+}=\frac{N}{2}+\frac{M}{2}$, where $M>0$ one gets (cf.~with Eq.~(\ref{masterEq12a}))
\begin{equation}\label{masterEq12b}
f\!\left(N_{\!_+}\right)= f\!\left(\frac{N}{2}\right)\left(\frac{1\!-\!q}{q}\right)^{\!\frac{M}{2}}\!\!.
\end{equation}
 
Summarizing, from Eqs.~(\ref{masterEq12a}) and~(\ref{masterEq12b}) it directly follows that, in the stationary regime of the majority-vote model on a complete graph with an even number of nodes, the probability of occurrence of a configuration $\Omega$ is given by (see Eq.~(\ref{assum0})): 
\begin{equation}\label{POmegaNeven}
P(\Omega)=P_0(q)\sqrt{\frac{1-q}{q}}^{\;|M(\Omega)|},
\end{equation}
where 
\begin{equation}\label{Pzero}
P_0(q)=f\!\left(\frac{N}{2}\right)
\end{equation}
is the probability of a microstate with zero magnetization, which can be calculated from the normalization condition: 
\begin{equation}\label{norma}
\sum_{\Omega}P(\Omega)=\sum_{N_{\!_+}=0}^N \binom{N}{N_{\!_+}} f(N_{\!_+})=P_0(q)\sum_{N_{\!_+}=0}^N \binom{N}{N_{\!_+}} \sqrt{\frac{1\!-\!q}{q}}^{\;|2N_{\!_+}-N|}=1.
\end{equation}
Let us also note that, in Eq.~(\ref{norma}), the expression under the second sum stands for the probability of the system to have exactly $N_{\!_+}$ positive spins, or to have magnetization $M=2N_{\!_+}\!-\!N$, i.e.
\begin{equation}\label{PMNeven}
P(M)=P_0(q) \binom{N}{\frac{N+M}{2}}\sqrt{\frac{1-q}{q}}^{\;|M|}.
\end{equation}

\begin{figure}[!t]
	\centering\includegraphics[width=0.7\columnwidth]{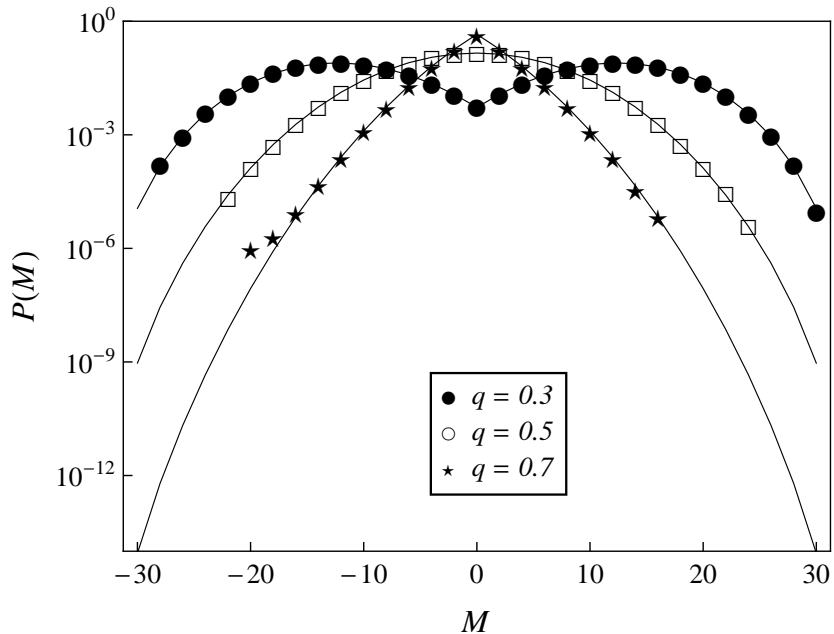}
	\caption{Probability $P(M)$ that the magnetization of the majority-vote model on a complete graph of size $N=30$ is equal to $M$. The scattered points  represent results of numerical simulations averaged over $10^4\times N$ independent realizations of the model. Different symbols correspond to different values of the noise parameter $q$, in accordance with the description given in the figure. The solid curves stand for the theoretical prediction according to Eq.~(\ref{PMNeven}).}
	\label{figPM}
\end{figure}

Finally, proceeding in a similar way as shown in Eqs.~(\ref{masterEq6a})-(\ref{PMNeven}), one can also find exact expressions for $P(\Omega)$ and $P(M)$ in systems with an odd number of nodes, $N$, and consequently, with an odd magnetization, $M$. The mentioned expressions have the following form:
\begin{equation}\label{POmegaNodd}
P(\Omega)=P_{_{+1}}(q)\sqrt{\frac{1-q}{q}}^{\;|M(\Omega)|-1},
\end{equation}
and 
\begin{equation}\label{PMNodd}
P(M)=P_{_{+1}}(q)\binom{N}{\frac{N+M}{2}}\sqrt{\frac{1-q}{q}}^{\;|M|-1},
\end{equation}
where $P_{_{+1}}(q)$ is the probability of a microstate with magnetization $M\!=\!+1$, and it can be shown that $P_{\!_{+1}}(q)=P_{\!_{-1}}(q)$.

\subsection{Comparison between theoretical predictions and numerical simulation results}\label{Computer}

\begin{figure}[!t]
	\centering\includegraphics[width=0.7\columnwidth]{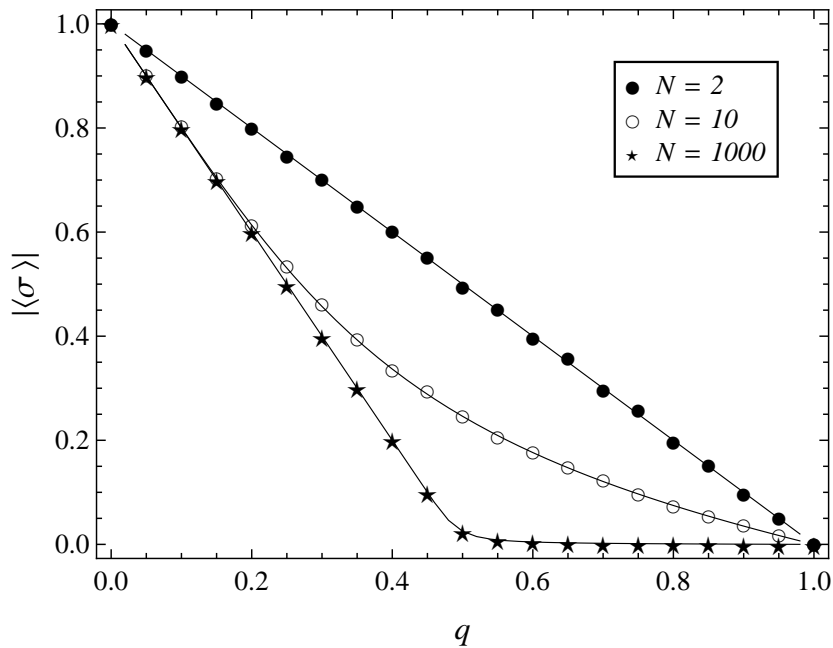}
	\caption{Absolute value of the average magnetization per spin, $|\langle\sigma\rangle|$, vs the noise parameter, $q$, in the MV model on complete graphs of various sizes, $N$. The scattered points represent results of numerical simulations for different values of $N$ (see description of the corresponding symbols in the figure). The solid curves result from the theoretical prediction: $|\langle\sigma\rangle|\!=\!\frac{\langle|M|\rangle}{N}$, where $\langle|M|\rangle\!=\!\sum_M|M\!|P(M)$, and $P(M)$ is given by Eq.~(\ref{PMNeven}).}
	\label{figsigma}
\end{figure}

The above theoretical predictions can be verified by numerical simulations. In particular, in Fig.~\ref{figPM}, theoretical and numerical probability distributions of magnetization, $P(M)$, for different values of the noise parameter, $q$, are shown to perfectly agree with each other. In this figure, for $q<\frac{1}{2}$, one can see that $P(M)$ is symmetric and bimodal. The behavior indicates that below the critical value of the noise parameter, $q_c=\frac{1}{2}$, the considered MV model is in a ferromagnetic phase. (The value of $q_c=\frac{1}{2}$ has also been recently obtained from mean-field analysis as the limiting case of the critical noise in classical random graphs; see~Eq.~(12) in~\cite{2015_PREChen}.) The ferromagnetic ordering for $q<q_c$ is characterized by a non-zero absolute value of the average magnetization per spin, see Fig.~\ref{figsigma}. 

\begin{figure}[!t]
	\centering\includegraphics[width=0.7\columnwidth]{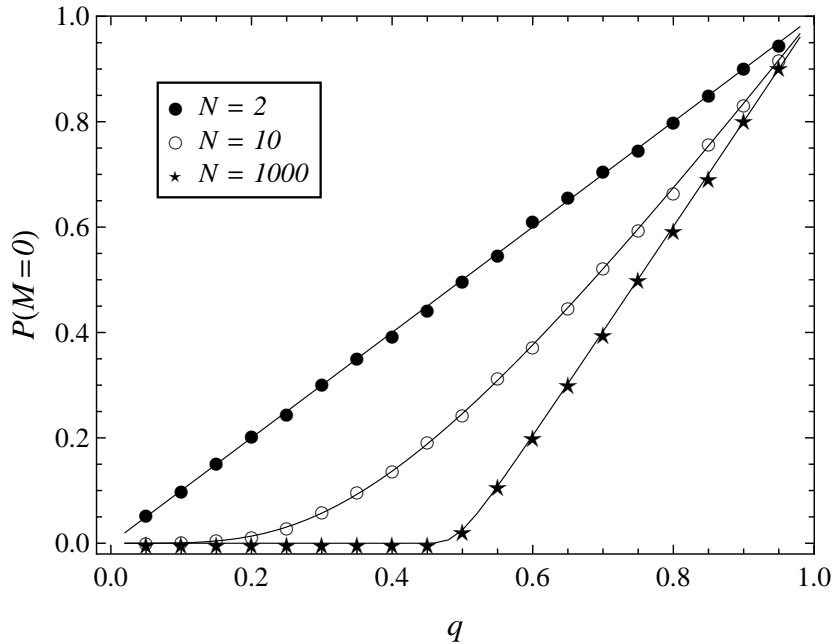}
	\caption{The probability that magnetization of the system is equal to zero, $P(M\!=\!0)$, vs the noise parameter, $q$, for different system sizes, $N$. As in the previous figures, the scattered points represent numerical simulation results, while the solid curves are the theoretical predictions according to Eq.~(\ref{PMNeven}), i.e. $P(M\!=\!0)=\binom{N}{\frac{N}{2}}P_0(q)$.}
	\label{figPM0}
\end{figure}

For $q>\frac{1}{2}$, the distribution $P(M)$ is unimodal with the most likely value of the magnetization equal to zero, see Fig.~\ref{figPM}. In this range of the noise parameter, i.e. above the critical value of $q_c=\frac{1}{2}$, the probability that magnetization of the system is equal to zero, $P(M\!=\!0)$, is a monotonically increasing function of $q$ (see Fig.~\ref{figPM0}). This probability reaches its maximum value of unity for $q\!=\!1$, when the system is an ideal antiferromagnet. (See discussion of the parameter $q$ below Eq.~(\ref{defmi}) in Sect.~\ref{SectBackground}.) 

\subsection{Detailed balance condition}\label{DB}

Knowing the exact expression for the probability distribution, $P(\Omega)$, one can show that the MV model on a complete graph fulfills the detailed balance condition (cf. with Eq.~(\ref{masterEq3})):
\begin{equation}\label{DB1}
\frac{P(\Omega)}{P(\Omega_i)}=\frac{w_i(\Omega_i)}{w_i(\Omega)}.
\end{equation}

Below, as in Sect.~\ref{Theory}, we consider in detail only the case of an even system size,~$N$. The case of an odd $N$ may be analyzed in a similar way. 

Thus, from Eq.~(\ref{POmegaNeven}) one gets that
\begin{equation}\label{DB2}
\frac{P(\Omega)}{P(\Omega_i)}=\sqrt{\frac{1-q}{q}}^{\;\Delta|M|},
\end{equation}
where
\begin{equation}\label{DB3}
\Delta|M|=|M(\Omega)|-|M(\Omega_i)|.
\end{equation}
It is easy to see that, since
\begin{equation}\label{DB4}
|M(\Omega_i)|=|M(\Omega)-2\sigma_i|=\left\{ \begin{array}{lcl}
|M|+2 & \mbox{for} & \sigma_iM\leq 0\\
|M|-2 & \mbox{for} & \sigma_iM\geq 2\\
\end{array}\right.,
\end{equation}
then the difference of the absolute values of magnetization in successive spin configurations is given by:
\begin{equation}\label{DB5}
\Delta|M|=\left\{ \begin{array}{lcl}
-2 & \mbox{for} & \sigma_iM\leq 0\\
+2 & \mbox{for} & \sigma_iM\geq 2\\
\end{array}\right..
\end{equation}
Therefore, the left hand side of the detailed balance condition, Eq.~(\ref{DB2}), becomes:
\begin{equation}\label{DB6}
\frac{P(\Omega)}{P(\Omega_i)}=\left\{ \begin{array}{lcl}
q/(1-q) & \mbox{for} & \sigma_iM\leq 0\\
(1-q)/q & \mbox{for} & \sigma_iM\geq 2\\
\end{array}\right..
\end{equation}

In a similar way, one can show that the right hand side of Eq.~(\ref{DB1}), which is a quotient of the rate transitions $w_i(\Omega)$ and $w_i(\Omega_i)$, also depends only on $\sigma_iM$. To see this, let us note that, in Eqs.~(\ref{rate1a}) and~(\ref{rate1b}), the product $\sigma_iS(M-\sigma_i)$ may have only two values:
\begin{equation}\label{DB7}
\sigma_iS(M\!-\!\sigma_i)=S\left(\sigma_i(M\!-\!\sigma_i)\right)= S(\sigma_iM\!-\!1)
=\left\{ \begin{array}{lcl}
\!-1 & \mbox{for} & \sigma_iM\leq 0\\
\!+1 & \mbox{for} & \sigma_iM\geq 2\\
\end{array}\right..
\end{equation}
Accordingly, the right hand side of Eq.~(\ref{DB1}) can be written as:
\begin{equation}\label{DB8}
\frac{w_i(\Omega_i)}{w_i(\Omega)}=\left\{ \begin{array}{lcl}
q/(1-q) & \mbox{for} & \sigma_iM\leq 0\\
(1-q)/q & \mbox{for} & \sigma_iM\geq 2\\
\end{array}\right..
\end{equation}

The correspondence between Eqs.~(\ref{DB6}) and~(\ref{DB8}) allows one to state that the detailed balance condition holds true in the majority-vote model on complete graphs. This means that the considered system is ergodic and, in the stationary regime, there exists its equilibrium representation in the sense of the canonical ensemble, 
\begin{equation}\label{PHOmega}
P(\Omega)\propto e^{-\mathcal{H}(\Omega)},
\end{equation}
with the Hamiltonian given by (cf.~Eqs.~(\ref{POmegaNeven}) and~(\ref{POmegaNodd})): 
\begin{equation}\label{HOmega}
\mathcal{H}(\Omega)=\ln\sqrt{\frac{q}{1-q}}|M(\Omega)|.
\end{equation}

\section{Summary}

The presented work is theoretical in nature. We have studied the isotropic majority-vote model which, apart from the one-dimensional case, is thought to be non-equilibrium. We found that if this model is defined on a complete graph, then, in the stationary regime, it has an equilibrium representation in the sense of the canonical ensemble. We show that the probability distribution, $P(\Omega)$, of finding the system in a certain microstate $\Omega=(\sigma_1,\sigma_2,\dots,\sigma_N)$, where $\sigma_i=\pm 1$, depends only on the absolute value of magnetization: $P(\Omega)\propto \sqrt{(1-q)/q}^{|M(\Omega)|}$, where $M(\Omega)=\sum_{i=1}^N\sigma_i$ and $q$ is the noise parameter of the model. Our theoretical predictions perfectly agree with the results of numerical simulations performed for systems of various sizes, $N\geq 2$. Analytical results, which are described in this work, are the first step to determine exact values of the mean-field critical exponents of the MV model, which are the subject of our next contribution.

\ack The authors wish to thank dr. hab. Andrzej Krawiecki for introducing them to the topic of the MV model. This work has been supported by the National Science Centre of Poland (Narodowe Centrum Nauki, NCN) under grant no.~2015/18/E/ST2/00560.

\end{document}